\documentstyle[pre,aps,psfig]{revtex}

\begin{document}

\draft
\twocolumn[\hsize\textwidth\columnwidth\hsize\csname@twocolumnfalse\endcsname

\title{Generalized Thermal Lattice Gases}

\author{O. Baran$^{\dagger}$, C. C. Wan$^{\dagger ,*}$ and
R. Harris$^{\dagger}$}

\address{${\dagger}$ Centre for the Physics of Materials, McGill
University, Montr\'eal, Qu\'ebec, Canada, H3A 2T8}

\address{$*$ Department of Physics, Fudan University, Shanghai 200433, China}

\date{\today}

\maketitle

\begin{abstract}

We show how to  employ thermal lattice gas models to describe 
non-equilibrium phenomena. This is achieved by
relaxing the restrictions of the usual micro-canonical
ensemble for these models via the  introduction of
thermal ``demons'' in the style of Creutz.  Within the Lattice
Boltzmann approximation, we then derive general expressions for the
usual transport coefficients of such models, in terms of the
derivatives of their equilibrium distribution functions.  To illustrate
potential applications, we choose a model obeying Maxwell-Boltzmann
statistics, and simulate Rayleigh-B\'enard convection with a forcing
term and a temperature gradient, both of which are continuously
variable.
\end{abstract}


\vskip1pc]
\narrowtext

\section{Introduction}

Frisch, Hasslacher and Pomeau (FHP) \cite{FHP,LGA}
pioneered the use of Lattice Gas Automata (LGA) to simulate the 
Navier-Stokes (NS) fluid. The motion of fictitious particles on an underlying
hexagonal
lattice, subject to carefully chosen rules for collisions and propagation,
gives rise to the NS equations in the continuum limit. 
Since that time, the LGA model 
and its derivative, the Lattice Boltzmann (LB) model \cite{LB1}, 
have attracted considerable  attention because of their 
potential application to the simulation of complex fluid 
systems, in particular, systems with inter-particle interactions which
model phase transitions and the dynamics of interfaces.  
Although the earliest such  models \cite{red+blue}
achieved spatial variation of the order parameter only 
by ignoring even semi-detailed balance in the description of possible 
collisions, more recent models are free from such inconsistencies
\cite{yeomans}.

However, like the original FHP model, all these models are intrinsically
``athermal'' \cite{ernst1}, and so are unable to simulate 
phenomena where the temperature is an  important variable.  Only recently have
thermal LGA models been constructed, such that the thermodynamics of fluids 
can be studied \cite{ernst2,thermal2,thermal3,chen,XWS}, and,
typically, they are defined within a micro-canonical ensemble.
To satisfy strict conservation rules for particle number, 
momentum and energy, they must contain several species of particles with
different energies, interacting via rules which are often very complex.
Reference \cite{thermal2}, for example, is an extension of the FHP model
to $4$ species of particles  with carefully chosen momenta and energies,
and in references \cite{thermal3,chen},  the authors have to introduce unequal
masses for particles moving in different channels, in order to ensure  a
sufficient number of allowed collisions.
Thermal LB models are in principle less complex because they are expressed
in terms of the distribution functions for the fictitious particles,
but have not been conspicuously successful \cite{check_these}.

A more serious problem for the study of thermal phenomena, particularly those
of a non-equilibrium nature
(heat conduction) or involving instabilities (convection),
arises because both LGA and LB thermal models
treat the temperature as an externally defined parameter.
However, when the temperature is itself a spatially varying parameter,
it is necessary to have information about its variation 
in order to implement the collision rules (LGA) or the
relaxation process (LB). Consequently, when this spatial variation is
itself the object of study, it is not possible without modification 
\cite{shan} to apply existing models.

Our purpose in this paper is to introduce a class of thermal models
which permit the study of non-equilibrium phenomena. As a by-product,
these models permit a much wider variety of collision rules for LGA
models, and are also readily adapted to the LB approach. The key
feature is a novel idea drawn from the Monte-Carlo literature
\cite{creutz1,creutz2,creutz3}.  In the language of the Ising model,
each site is associated with a local thermal reservoir or ``demon'',
which interchanges energy with the particles on the site.  The local
thermodynamic temperature, in the low velocity limit, is then
proportional to the local demon energy.  In the LGA or LB models, each
demon is associated with one node of the hexagonal lattice, and
constitutes a mechanism for monitoring or controlling the {\em local}
temperature, so that the modeling of non-equilibrium phenomena becomes
possible.

In the following section, we describe the equilibrium properties of
such thermal lattice gas models. Then, in Section~\ref{sec:transport},
we derive expressions for the transport coefficients of Lattice
Boltzmann models by means of a Chapman-Enskog expansion.  In
Section~\ref{sec:amodel}, we illustrate ideas using a particular
model, and in Section~\ref{sec:convection}, as a specific example, use
this model to simulate two-dimensional Rayleigh-B\'enard convection
and to draw some preliminary conclusions concerning the feasibility of
our approach.

\section{Thermal Lattice Gas Models}
\label{sec:general}
\subsection{Generalities}
In analogy with the basic LGA model \cite{LGA}, we define a class of
thermal models in which several species of
fictitious particles move on a two-dimensional
hexagonal lattice.  Some of the particles (rest particles) remain at
rest.  During each time step, the particles interact (``collide'' like
billiard balls) and then propagate ballistically at constant speed from
one site of the lattice to a neighbouring site.  The collision step
rigorously conserves the number of particles and their total momentum.
Energy is also conserved, but in a particular manner, described below.

Energy is defined as the sum of the kinetic and potential energies of the
individual fictitious particles, and it is essential that there be at least two
species of particles
with distinct energies for there to be well-defined thermal properties.
In the literature, such thermal models often
employ only kinetic energies, so that within the micro-canonical ensemble,
particles with different speeds
and or masses \cite{thermal2,thermal3,chen} are required to satisfy the
conservation laws.
(This restriction to kinetic energies and to collision processes which
strictly conserve energy also has
the consequence \cite{ernst3} that such models posses zero bulk viscosity.)
However, the inclusion of potential energy terms is straightforward, 
\cite{chen,ernst2}, although 
seldom employed for other than pedagogical purposes.

Conservation of energy is enforced with the aid of the so-called demons, one at
each site. The advantage of this procedure is that it is not necessary to
choose values of masses, speeds and/or energies to permit a sufficient
number of non-trivial collisions. Instead, a demon acts as a kind of energy
reservoir, permitting collisions in a manner which satisfies detailed balance.
Over the course of time, there will be a distribution
function for the local demon energy, which has the form
$P(E_D) \sim \exp{-\beta E_D}$, where $\beta = 1/T$ 
is the inverse
of the local temperature. Since $E_D$ is a continuously variable quantity,
its average value $<\!E_D\!>$ is equal to the local temperature.
The idea is borrowed from Creutz \cite{creutz1}, and was exploited
in a series of papers applying Ising-style lattice gas models to
non-equilibrium interface problems \cite{loki}. 

At each site, a collision process may proceed {\em either} if it produces
surplus (or zero)
energy {\em or} if the local demon can provide the requisite energy
deficit. In the former case, the demon absorbs the surplus. Demons are
required always  to have positive energy, and as a result serve also to
regulate the occurrence of collision processes. As demonstrated by Creutz
and by J\"orgenson {\it et al} \cite{loki}, the demons thus act as 
thermometers, measuring
the local temperature by virtue of their own statistically averaged energy.
It is also possible to use the demons to control the local temperature: if
this is done at the boundaries of a sample, then, for example, a temperature
gradient can be set up across the sample, permitting a measurement of thermal
conductivity \cite{hg} or the establishing of Rayleigh-B\'enard convection.

A useful way to understand the role of the demons is to consider them as a
species of rest particle. Rest particles of the conventional kind can be 
created or annihilated in collision processes so as to satisfy the
conservation laws. Demons  are neither created nor annihilated, but act to
satisfy the conservation of energy. Conventional rest-particles (usually) have
only one energy level, but demons have many such levels - although in simple
situations one might imagine demons with only a small number of distinct
levels \cite{creutz3}.

\subsection{Statistical Equilibrium}

As a first step in deriving expressions for the transport coefficients 
we define some generic notation \cite{LGA}.
Sites in the hexagonal lattice are labeled by an index {\it
i} and by a site vector ${\bf R}_i$.
The vectors radiating from a site {\it i}
to its (not-necessarily) nearest neighbours define different possible 
directions for the motion of fictitious particles at that site.
For particles of species $I$, we have
${\bf c}_{a,I}, ~a=1,2,\cdots,b_I$, where $b_I$ is the number of distinct
neighbours at distance $c_I$.  This distance also defines the ``speed'' of the
species $I$, so that the kinetic energy of
each fictitious moving particle is $\frac{1}{2} c_I^2$. 
In similar fashion, each particle of species $I$
has potential energy $\epsilon_I$,
so that its total energy  is $E_I = \frac{1}{2} c_I^2 + \epsilon_I$.
Any rest particle has total energy zero.


The fictitious particles occupy the available states $\{i,a,I\}$
according to particular statistics,
and the detailed balance property  of the collision rules guarantees that the
system possesses a state of statistical equilibrium.
Furthermore, the existence of an equilibrium distribution 
function is an essential
condition for the regaining of the Navier-Stokes equations in the continuum
limit \cite{LGA}. The usual choice of statistics is Fermi-Dirac, since,
historically, it was convenient for coding purposes to represent occupation
numbers as binary variables. However, this choice is not essential. 
In general we write the occupation number of the state $\{i,a,I\}$ as
$n_{i,a,I}$, and the ensemble-averaged distribution function for this state as
$f_{i,a,I} = <\!n_{i,a,I}\!>$. It is convenient to represent rest particles in the
same way by $n_{i,0}$ and $f_{i,0}$.

In an equilibrium state for which there is no net motion of the fluid,
the functions $f$ have the values $f^{eq}_I$, so that
the probability of finding a certain configuration can then be written 
as a product of the continuous variables $f^{eq}$ and $P(E_D)$, the latter 
variable being the probability of finding a demon with energy $E_D$. 
It is now possible to define thermodynamic
variables in terms of the equilibrium distributions.

The most important of these are the density $\rho$ and the
pressure, $P$. The density is just
\begin{equation}
 \rho = M f^{eq}_0 + \sum_{a,I} f^{eq}_I \equiv \rho_0 + \sum_I \rho_I
\label{density}
\end{equation}
where $M$ is the number of rest-particle states per site.
In principle, the pressure
should be derived from the free energy. However, as is well known
in the lattice gas literature \cite{LGA,ernst1}, the quantity which plays
the role of pressure in the hydrodynamic equations is 
\begin{equation}
P = \frac{1}{2} \sum_I c_I^2 \rho_I 
\label{pressure}
\end{equation}
sometimes known as the ``kinetic pressure''. It is useful to define the
analogous partial pressures
$ P_I = \frac{1}{2} c_I^2 \rho_I $, so that $P=\sum_I P_I$.
We will also require the energy density $U = \sum_I E_I \rho_I$
\cite{density}, and the
isothermal and adiabatic speeds of sound, which are respectively
\begin{equation}
c_T^2 = \frac{1}{2} \sum_I \rho_I c_I^2 /\rho 
\label{iso}
\end{equation}
and
\begin{equation}
c_s^2 = \frac{1}{2} \sum_I  f^{\mu}_I c_I^4 / \sum_I  f^{\mu}_I c_I^2 
\label{adiabat}
\end{equation}
The derivatives  $f^{\mu}_I$ of the equilibrium distribution functions
$f^{eq}_I$  are 
defined as 
$f^{\mu}_I = T \left(\frac{\partial f_I}{\partial \mu}\right)_{\!T}$
where $T$ is the temperature and $\mu$ is the chemical potential
\cite{note1}.

When there exists a net local flow velocity ${\bf u}$ 
it is still possible to define equilibrium distribution functions.
In the low velocity limit $u \ll c$, 
they  can be expanded
in powers of ${\bf u}$. Following reference \cite{LGA}, but with a slightly
generalised notation,  we obtain to first order \cite{note2}

\begin{eqnarray}
f^{eq}_{a,I}({\bf u}) = f^{eq}_{I} +
q \rho f^{\mu}_I {\bf c}_{a,I} \cdot {\bf u} 
\label{taylor}
\end{eqnarray}

The constant $q$ is determined by requiring that the expansions
be consistent with the total momentum defined as
\begin{equation}
\rho {\bf u} = \sum_{a,I} {\bf c}_{a,I} f^{eq}_{a,I} 
\label{velocity}
\end{equation}
We obtain
\begin{eqnarray}
q=  2 / \sum_I b_I f^{\mu}_I c_I^2 \\
\label{q}
\end{eqnarray}
and it is convenient to define $q_I = q  f^{\mu}_I$ so that 
\begin{eqnarray}
\label{f_eq}
f^{eq}_{a,I}({\bf u}) = f^{eq}_I +
\rho q_I {\bf c}_{a,I} \cdot {\bf u} 
\end{eqnarray}
and $\frac{1}{2} \sum b_I q_I c_I^2 = 1$.
Note that to first order in ${\bf u}$,
there are no corrections to $f^{eq}_0$ or to $P(E_D)$.

\subsection{Time Evolution}

As in a traditional LGA \cite{LGA}, time evolution  proceeds in two distinct
steps. First, particles at a given site,
(both rest particles and moving particles in any energy level),
interact with each other (``collide'') following predetermined rules \cite{LGA}.
Energy conservation is ensured by the local demon, as described previously,
and detailed balance is rigorously observed.
Conservation of particles, momentum and energy on each site
can be written explicitly as
\begin{equation}
\sum_{a,I} f_{i,a,I}(t) + f_{i,0}(t) = {\textstyle constant}
\label{eq.2.1}
\end{equation}
\begin{equation}
\sum_{a,I} {\bf c}_a f_{i,a,I}(t) = {\textstyle constant}
\label{eq.2.2}
\end{equation}
\begin{equation}
\sum_{a,I} f_{i,a,I}(t)E_I + E_{D,i}(t) =  {\textstyle constant}
 \label{eq.2.3}
\end{equation}
where $t$ is the (discrete) time. $E_{D,i}$ is the demon energy.

After each collision step, there is a propagation 
step, in which each moving particle moves one lattice constant in 
the direction of its velocity, while the rest particles and the demon remain 
unchanged. Thus the kinetic equations for the particle and demon variables,
including both collisions and propagation, are:
\begin{eqnarray}
f_{j,a,I}(t+1) - f_{i,a,I}(t) &=& \Omega_I,~~~a=1,\cdots,b_I \\ 
f_{i,0}(t+1)-f_{i,0}(t)&=&\Omega_0,  \nonumber \\
E_{D,i}(t+1)-E_{D,i}(t)&=&-\sum_{I}\sum_a^{b_I} \Omega_I E_I
\label{boltzmann}
\end{eqnarray}
where the index $j$ is defined by ${\bf R}_j={\bf R}_i + \hat{\bf c_a}$.
$\Omega$ is the ensemble averaged collision operator.
Strictly speaking, $E_{D,i}$ should also be replaced by its ensemble average,
$<\!E_{D,i}\!>$.

The lattice Boltzmann approximation to  the collision matrix directly employs
the existence of equilibrium distribution functions. Any distribution function
which deviates from its equilibrium value can be written as $f = f^{eq} +g$,
where $g$ is hopefully small. If we assume that there exists a single
characteristic relaxation time $\tau$ \cite{BGK}, 
then in terms of $g$, we may approximate the rate equations as 

\begin{eqnarray}
f_{j,a,I}(t+1) - f_{i,a,I}(t) &=& -g_{i,a,I}/\tau, ~~~ a=1,\cdots,b_I 
\label{lboltzmann} \\
f_{i,0}(t+1) - f_{i,0}(t) &=& -g_{i,0}/\tau \label{lboltz0}
\end{eqnarray}
These  expressions will be employed to
simplify the analysis in the subsequent parts of the paper.

\section{Transport Coefficients}
\label{sec:transport}

The basic continuum equations for a thermal fluid could now be obtained 
via a Chapman-Enskog expansion. However, since such treatments are available
elsewhere \cite{chen}, we will focus only on the derivation of expressions 
for the transport coefficients. We will follow the procedure of reference
\cite{LGA}. The first step is to  replace
the rate equations (\ref{lboltzmann}), (\ref{lboltz0}) by their continuum
versions. In the limit where spatial and temporal changes are small and/or
slow, the left-hand sides of these equations yield derivatives which
can be replaced by $\partial_t=\epsilon\partial_{t_1}$
$+\epsilon^2\partial_{t_2}$, $\nabla{r}=\epsilon\nabla{r_1}$, where
$\epsilon$ is a small parameter.
The distribution functions can also be expanded in terms of $\epsilon$. 
To order $\epsilon^2$, this gives: 
\[f_{a,I}= f^{(0)}_{a,I}+\epsilon f^{(1)}_{a,I} +\epsilon^2 f^{(2)}_{a,I}, \]
and
\[f_{0,I}= f^{(0)}_{0,I}+\epsilon f^{(1)}_{0,I} +\epsilon^2 f^{(2)}_{0,I} \]
where the zeroth order terms in the expansion are just
the equilibrium distributions, and the higher order
terms are just $g$: $f^{eq} \equiv f^{(0)}$ and 
$g \equiv \epsilon f^{(1)} + \epsilon^2 f^{(2)}$. 
Necessarily, the conservation of mass, momentum and energy
require that 
\begin{eqnarray}
\label{cons_m}
f^{(\alpha)}_0+\sum_{a,I}f^{(\alpha)}_{a,I}=0,  \\
\label{cons_p}
\sum_{a,I}{\bf c}_a f^{(\alpha)}_{a,I}=0
\end{eqnarray}
and
\begin{eqnarray}
\label{cons_e}
E_D^{(\alpha)}+\sum_{a,I}E_I f^{(\alpha)}_{a,I}=0
\end{eqnarray}
where $\alpha=1,2$. 

Substituting into equations (\ref{lboltzmann})-(\ref{lboltz0}), 
and equating terms for each
order of $\epsilon$, we obtain a hierarchy of Boltzmann
equations. The first order equations are: 
\begin{eqnarray}
\label{first}
D^{(1)}_{a,I}  f^{(0)}_{a,I} \equiv (\partial_{t_1}+
{\bf c}_{a,I} \cdot\nabla_1)
f^{(0)}_{a,I}= -{1\over\tau}f^{(1)}_{a,I}, \\
\partial_{t_1} f^{(0)}_0=-{1\over\tau} f^{(1)}_0,\\
\partial_{t_1}E_D^{(0)}=-{1\over\tau}E_D^{(1)}.
\end{eqnarray}

Using (\ref{cons_m}) and (\ref{cons_p}) we then obtain
\begin{eqnarray}
\partial_{t1} (\rho) + \nabla_1 \cdot (\rho {\bf u}) = 0
\label{mass1}
\end{eqnarray}
and 
\begin{eqnarray}
\partial_{t1} (\rho {\bf u}) + \nabla_1 P = 0
\label{momentum1}
\end{eqnarray}
where $P$ is the kinetic pressure as defined earlier.
Similarly, since these relations are true for any values of the
distribution functions $f^{eq}({\bf u}=0)$, there must also be analogous
relations for the partial densities $\rho_I$ and partial pressures $P_I$,
namely
\begin{eqnarray}
\partial_{t1} (\rho_I) + \frac{\rho_I}{\rho} \nabla_1 \cdot (\rho {\bf u}) = 0
\label{pmass1}
\end{eqnarray}
and
\begin{eqnarray}
\partial_{t1} (\rho_I {\bf u}) + \nabla_1 P_I = 0
\label{pmomentum1}
\end{eqnarray}

The expressions for the shear viscosity $\nu$, the bulk viscosity $\zeta$  
and the thermal conduction coefficient $\lambda$ arise from the second order
terms  in $\epsilon$. After some reduction, the corresponding equations become
\begin{eqnarray}
\partial_{t2} (\rho) + \frac{1}{2} \sum_{a,I} D^{(1)}_{a,I} D^{(1)}_{a,I}
 f^{(0)}_{a,I} =0
\label{mass2}
\end{eqnarray}
and
\begin{eqnarray}
\partial_{t2} (\rho {\bf u}) &+& \frac{1}{2} \sum_{a,I} D^{(1)}_{a,I} 
D^{(1)}_{a,I} {\bf c_{a,I}} f^{(0)}_{a,I} \nonumber \\
&+& \sum_{a,I}  {\bf c_{a,I}}{\bf c_{a,I}} \cdot \nabla_1  f^{(1)}_{a,I} =0
\label{momentum2}
\end{eqnarray}
where the terms preceded by the factor $\frac{1}{2}$ are second order in
the Taylor expansion of the derivatives.
These terms can be reduced by making use of the
equations (\ref{cons_m}) and (\ref{first}).
Although equation (\ref{mass2}) reduces to the simple result 
$\partial_{t2} (\rho) =0$, 
equation (\ref{momentum2}) is more interesting. 
We simplify its second term
to read  $ \sum_{a,I} {\bf c_{a,I}} \cdot \nabla_1 D^{(1)}_{a,I} 
{\bf c_{a,I}} f^{(0)}_{a,I}$,
and then substitute for $f^{(1)}_{a,I}$ from (\ref{first}) to obtain
\begin{eqnarray}
\partial_{t2} \rho ({\bf u}) = (\tau - \frac{1}{2}) \sum_{a,I} {\bf c}_{a,I}
\cdot \nabla_1 D^{(1)}_{a,I} {\bf c_{a,I}} f^{(0)}_{a,I}
\label{almost}
\end{eqnarray}
The final step is to substitute (\ref{f_eq}), and use (\ref{pmass1}) and
(\ref{pmomentum1}) to write
\begin{eqnarray}
\partial_{t2} (\rho {\bf u}) = (\tau - \frac{1}{2}) 
\sum_I ( X_I \nabla^2 \rho  {\bf u} + Y_I \nabla \nabla \cdot \rho {\bf u})
\label{almost2}
\end{eqnarray}
where
\[ X_I = \frac{1}{4} b_I q_I c_I^4 \]
and
\[ Y_I = \frac{1}{2} (b_I q_I c_I^4 - \rho_I c_I^2 / \rho ) \]

Combining with (\ref{momentum1}), this becomes
\begin{equation}
\partial_t (\rho {\bf u}) = -\nabla P + (\tau - \frac{1}{2}) (\nu \nabla^2 
\rho {\bf u} + \zeta \nabla \nabla \cdot \rho {\bf u} )
\end{equation}
where the shear viscosity  $\nu$ is 
\[\nu = \frac{1}{2}c_s^2 (\tau- \frac{1}{2}) \]
and the bulk viscosity $\zeta$ is 
\[ \zeta =(c_s^2 - c_T^2)(\tau- \frac{1}{2})\]
using the expressions for the sound speeds $c_T$ and $c_s$
given earlier.
Note that the bulk viscosity vanishes for any model having only one value 
of the speeds $c_I$: this includes any FHP1 model, \cite{LGA},  
as a special case.

The procedure to obtain the thermal conduction coefficient, $\lambda$, is 
very similar to that described, for example, in Huang \cite{huang}.
A key feature is the necessity to subtract out of the energy current that part
which depends on the net flow of the fluid.
This is the origin of the ``subtracted current'' described by Ernst
\cite{ernst2,ernst3}, but neglected by Chen et al \cite{chen}.
We account for this effect by replacing  the energies $E_I$ by $\tilde{E}_I$
such that
\begin{equation}
\sum_{a,I} {\bf c}_{a,I} \tilde{E}_I f^{eq}_{a,I}({\bf u}) =  0
\end{equation}
or
\begin{equation}
\sum_I b_I f_I^{\mu} c_I^2 \tilde{E}_I  =  0
\label{tilde_e}
\end{equation}
Writing $\tilde{E}_I = E_I - E_S$, we obtain
\begin{eqnarray}
E_S &=& \sum_I b_I q_I E_I c_I^2 / \sum_I b_I q_I c_I^2 \nonumber \\
    &=& \sum_I b_I f^{\mu}_I E_I c_I^2 / \sum_I b_I f^{\mu}_I c_I^2
\end{eqnarray}

The first order equation describing the  time evolution of the
local energy density $\tilde{U}= \sum_{a,I} f^{eq}_{a,I}({\bf u}) \tilde{E}_I$
is then
\begin{eqnarray}
\partial_{t1}  [ \tilde{U} + E_D ] = 0
\end{eqnarray}
which gives no new information. However, 
after using equations (\ref{cons_e})-(\ref{first}),
the second order equation  becomes
\begin{eqnarray}
\partial_{t2} [ \tilde{U} + E_D ] = (\tau -\frac{1}{2})
\sum_{a,I} \tilde{E}_I {\bf c}_{a,I} \cdot \nabla D^{(1)}_{I,a}
f^{eq}_{I,a}({\bf u}) 
\label{E2}
\end{eqnarray}
Using the explicit form of $f^{eq}_{I,a}({\bf u})$, equation (\ref{f_eq}),
and the definition of $\tilde{E}$, equation (\ref{tilde_e}), we therefore obtain
\begin{eqnarray}
\partial_t [\tilde{U} + E_D] = (\tau - \frac{1}{2})\sum_I \tilde{E}_I c_I^2 \nabla^2 \rho_I 
\end{eqnarray}
It remains to  express the gradient explicitly in terms of the thermal gradient.
In the absence of a net flow of particles, the chemical potential $\mu$ is
not a function of position, and so
$\nabla^2 f_I= f_I^T \nabla^2 T$, where $f_I^T =
\left(\frac{\partial f_I}{\partial T}\right )_{\!\mu} $,
and the coefficient of thermal conduction is
\begin{equation}
\lambda =(\tau -\frac{1}{2})  \sum_I b_I f_I^T c_I^2 \tilde{E}_I /2
\end{equation}
It is convenient to write this also in the form
\begin{equation}
\lambda =(\tau -\frac{1}{2})\frac{1}{T^2}  \sum_I b_I f_I^\beta c_I^2 \tilde{E}_I/2
\end{equation}
where $\beta = \frac{1}{T}$ and $f_I^\beta$ is given by 
$f_I^\beta = -\left(\frac{\partial
f_I}{\partial \beta}\right )_{\!\mu} $. The high temperature behaviour of
$\lambda$ is dominated by the $\frac{1}{T^2}$ prefactor.

For the particular case of the Fermi-Dirac distribution, it is easy to show
that this is precisely the expression given by Ernst \cite{ernst3}.
We write
$ f_I^\beta = - (\mu - E_I) f_I^{\mu} $
and therefore obtain
\begin{eqnarray}
\lambda &=&(\tau -\frac{1}{2}) \frac{1}{T^2} \sum_I b_I f_I^{\mu}
c_I^2 \tilde{E}_I (E_I - \mu)/2  \nonumber \\
        &\equiv& (\tau -\frac{1}{2}) \frac{1}{T^2} \sum_I b_I f_I^{\mu} c_I^2
\tilde{ E}_I^2/2
\end{eqnarray}
where the last step comes from the identity (\ref{tilde_e}).

\section{A Model}
\label{sec:amodel}

To illustrate the ideas of the previous sections, and, in particular, to
illustrate the use of statistics other than Fermi-Dirac, 
we have previously described a model where the statistics are quite
unconventional \cite{ccw}. In the present paper, we choose to employ
Maxwell-Boltzmann statistics for a
simple model with 3 energy levels. The lowest level,
with energy $E_0 = 0$, is $M$-fold degenerate,
so that there are at most $M$ fictitious
particles with zero velocity (``rest particles'').
The other two levels both correspond to the same speed $c$, 
with the same 6-fold symmetry as in the FHP models, but their
energies are respectively $E_A $ and $E_B = E_A + \Delta$. 
In our numerical calculations, we take $E_A = 0.62$ 
and $E_B = 1.80$ in units of $c^2$.

Assuming Maxwell-Boltzmann statistics, we obtain

\[ f_0({\bf u}) = e^{\beta \mu}; \,\,\, f_{I,a}({\bf u}) = e^{\beta(\mu - E_I)}
( 1 + q \rho {\bf u}\cdot c_a), \,\,\, I=A,B \]
where $q=\rho e^{-\beta \mu}/3 c^2 (e^{-\beta E_A} + e^{-\beta E_B})$,
and $\rho e^{-\beta \mu} = [M + 6 ( e^{-\beta E_A} + e^{-\beta E_B}) ]$.
This simple form for the velocity expansion results because
$f_I^{\mu} = f_I^{eq}$.

Because there is only one speed in the model, the expressions for the
sound velocities and for the shear and bulk viscosities become
particularly simple. We obtain
\begin{eqnarray}
c_T^2 &=& \frac{1}{2} c^2 \frac{\sum_{A,B} \rho_I }{ \rho}
= 3 c^2 \left [ \frac{e ^{-\beta E_A} + e^{-\beta E_B}}
{ M + 6(e ^{-\beta E_A} + e^{-\beta E_B})} \right ];  \nonumber \\
c_s^2 &=& \frac{1}{2} c^2
\end{eqnarray}
and
\begin{eqnarray}
\nu &=& (\tau -\frac{1}{2})c^2 /4;   \nonumber \\
\zeta &=& (\tau -\frac{1}{2})
\frac{c^2}{2} \frac{\rho_0}{\rho} \nonumber \\
&=& (\tau -\frac{1}{2})\frac{c^2}{2}
 \frac{M}{M + 6(e ^{-\beta E_A} + e^{-\beta E_B})}
\end{eqnarray}

The temperature-independence of $\nu$ is an artifact of the model, but the
temperature dependence of $\zeta$ is typical of the viscosity of
a liquid. 
For a typical value of the density, 
$\zeta$ is displayed in Figure 1 as a function of temperature.

The model also leads to a simple
expression for the thermal conduction coefficient $\lambda$:
\[ \lambda = (\tau -\frac{1}{2})\frac{3 c^2 \Delta^2}{T^2} \frac{f_A^{eq}
f_B^{eq}}{f_A^{eq}+f_B^{eq}} \]
For the purpose of illustration, it is convenient to relate this
to the thermal diffusion coefficient, defined as $D_T= \lambda/\rho C_p$,
which is  also plotted in Figure 1.
Although the temperature dependence of $\lambda$ is dominated by the 
$\frac{1}{T^2}$ prefactor at high temperatures, $T > \Delta$, this behaviour
is exactly compensated by the temperature dependence of the specific heat.

\begin{figure}
\hspace*{-1.0cm}
\centerline{\psfig{figure=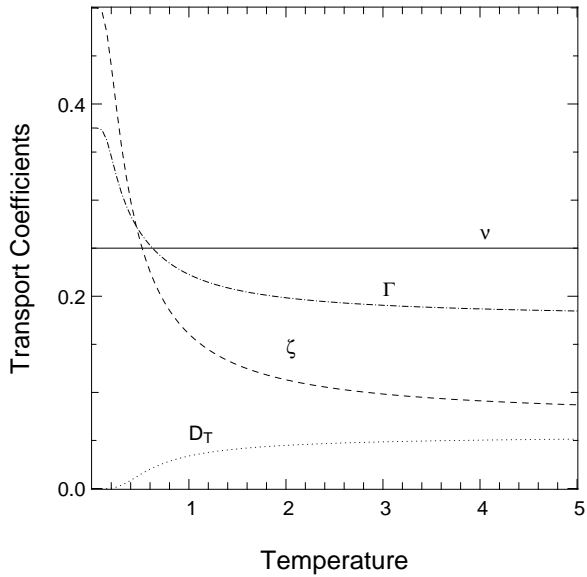,width=8.5cm}}
\caption{
Temperature variation of the transport coefficients for the model
described in the text.  Solid line: bulk viscosity. Dashed line: shear
viscosity.  Dotted line: thermal diffusion. Dash-dotted line: sound
attenuation.  All coefficients are in units of $\tau - \frac{1}{2}$.
}
\label{fig:transport}
\end{figure}

For completeness, Figure~\ref{fig:transport} also shows the
temperature variation of the sound attenuation coefficient $\Gamma$,
defined in the standard way as $\Gamma = \frac{1}{2}(\nu + \zeta +
(\gamma - 1) D_T)$.


\section{Rayleigh B\'enard Convection}
\label{sec:convection}


To illustrate the feasibility of our approach,
we carried out  simulations of Rayleigh-B\'enard convection.
We considered a two dimensional cell with horizontal length
$L$  and vertical height $H$. 
Periodic boundary conditions were
imposed in the horizontal direction, with two 
rigid walls at the top and the bottom of the cell. 
The demon energies on the upper and lower walls were fixed at
values $T_1$ and $T_2$, respectively. A uniform force was 
implemented by changing the vertical momentum of the  
particles at a constant rate, while keeping horizontal momentum unchanged. 
Thus, both the temperature difference 
$\delta T$ and the force could be tuned continuously. 

Our initial study was of
a system with $L = 400$ lattice units and $H=100 \times \sqrt{3}$ 
units, so that the aspect ratio $L : H$ was near $2 : 1$,
with a particle density of $3.6$ per site. 
The energy levels were $E_A=0.62$, $E_B=1.8$ in units of $c^2$,
and the temperatures at
the lower and upper boundaries were $T_1=4.8$, $T_2=0.3$ in the same units.
We chose the relaxation time $\tau$ to be 1 unit, and averaged velocity
fields over $20\times 20$ regions in order to evaluate the local
distribution functions.
Initial conditions were a uniform density and a uniform temperature
gradient, and with the local velocity everywhere zero,
so that we were able to observe the onset and subsequent
evolution of the convective instability.
With these parameters, the evolution of the system was sufficiently slow
that ``snapshots'' of the velocity field could be obtained by time averages
over only $50$ time-steps.

\begin{figure}
\hspace*{-0.5cm}
\centerline{\psfig{figure=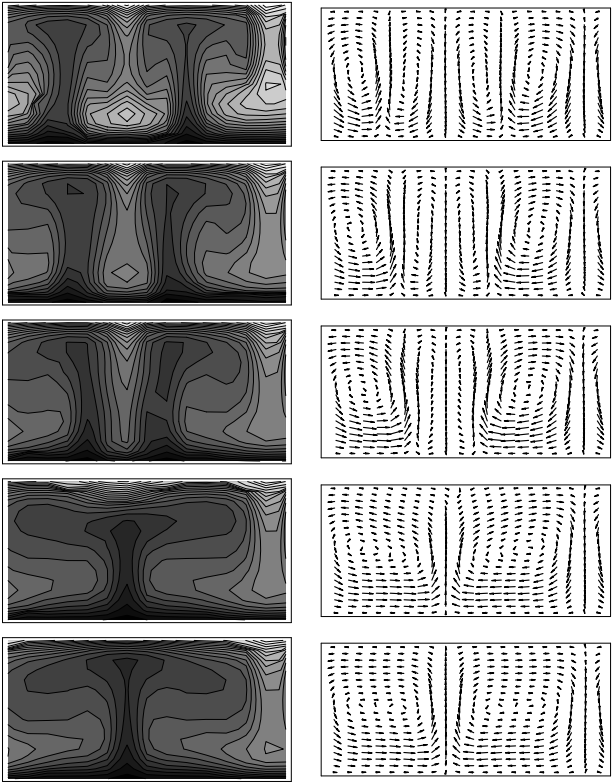,width=8.5cm}}
\vspace*{0.2cm}
\caption{Contour plots showing the time evolution 
of the temperature distribution (left panels) and
of the velocity field (right panels). From top to bottom, data is shown at
times 20,000, 40,000, 60,000, 80,000 and 100,000 units respectively.
One time unit is defined as one update of every site in the lattice.}
\label{fig:fieldplot}
\end{figure}

Figure~\ref{fig:fieldplot} shows the evolution of the temperature
distribution, and of the corresponding velocity fields for these
parameters.  At first four convection rolls are clearly seen, but
subsequently these collapse into two.  Evidently, the model has
captured the essential features of the physical phenomenon.
Systematic analysis of the data as a function of the system parameters
will be the subject of a subsequent publication, but certain
preliminary remarks are appropriate here.

According to the classic linear stability analysis, \cite{chandra},
convection occurs when the Rayleigh number $\cal{R}$ \cite{rayleigh}
exceeds a critical value of order $10^3$.  Even for situations far
from the linear regime, the value of $\cal{R}$ is a good indicator of
the stability of the convective phenomenon.  Thus, since $\cal{R}$ is
of order $10^{5}$ for the simulation shown in the figure, the
convection rolls should be extremely stable, as is indeed the
case. Indeed, our results are strikingly similar to those of a
previous detailed study \cite{shan}, obtained with a modified LB model
which represented the temperature as a ``convected passive scalar
field''.


An interesting feature of our results
is the observation of four rolls before the final stable
state with two rolls is reached. Linear analysis predicts two rolls (since
the wavenumber of the instability should be around $\pi/H$ where $H$ is the
height of the cell). Our tentative explanation is that 
the velocity field is first established near the lower (hot) boundary,
so that the effective height of the cell is considerably smaller than $H$.
We therefore expect  the wavenumber of the instability to be larger than
$\pi/H$, and the periodic boundary conditions select $\sim 2\pi/H$.
(In principle, with different geometry, we might expect to see yet higher
initial wavenumbers.) The study of this transient regime, and the subsequent
stabilisation of the two rolls, will form part of our ongoing
investigations.

However, in general terms, this preliminary illustration of
Rayleigh-B\'enard convection demonstrates the feasibility of
our Lattice Boltzmann method for non-equilibrium
thermal lattice-gases. We intend to apply
our approach to a variety of phenomena for which the statistical mechanics
of the gas are critically important. In particular, we plan to include
interactions between fictitious particles so as to simulate systems with
first order
phase transitions and the dynamics of interfaces between their associated 
phases.

We thank Hong Guo and Martin Grant for many useful discussions.

\noindent


\begin{references}
\bibitem{density}
Strictly speaking, the energy per unit area (rather than per site) requires
an additional factor of $\sqrt{3}/2$. This same factor would appear in the
expressions for the thermal conductivity and for the specific heats.
\bibitem {note1}
This is not the definition employed in
reference \cite{LGA}, although the  change in the algebra  is minimal.
The reason for the change is that it
is useful to introduce the thermodynamic variables appropriate to thermal
LGA models \cite{ernst3}. Because the volume of the system is by definition
fixed, it is the density which provides the link with continuum models. This
suggests that the notation of the grand canonical ensemble is the  most
appropriate, so that we will employ the temperature $T$ and the 
chemical potential $\mu$. 
\bibitem{rayleigh}
The Rayleigh number is conventionally defined as ${\cal{R}} =
\alpha_P g H^3\delta T \rho C_v/(\nu\lambda)$, where $\alpha_P$
is the thermal expansion coefficient, $g$ is the gravitational field,
$H$ is the height of the cell, and $\delta T$ is the temperature difference
between the two boundaries.
\end{references}
\end{document}